\documentclass[preprint,showpacs,preprintnumbers,amsmath,amssymb]{revtex4}

\usepackage{graphicx}% Include figure files
\usepackage{dcolumn}% Align table columns on decimal point   
\usepackage{bm}% bold math

\begin{document}

\preprint{YGHP-03-27}

\vspace *{0.5cm}
                       
\title{
Quiescent String Dominance Parameter F and Classification of One-Dimensional
Cellular Automata}

\author{Sunao Sakai}
\author{Megumi Kanno}
\author{Yukari Saito}
\affiliation{Faculty of Education, Yamagata University,
 Yamagata, Japan.}                                                  
 \email{sakai@e.yamagata-u.ac.jp}
% \altaffiliation[Also at ]{Faculty of Education, Yamagata University,
% Yamagata, Japan.}

\affiliation{%
}%

\date{\today}% It is always \today, today,

\begin{abstract}
The mechanism which discriminates the pattern classes at the same
$\lambda$, is found. It is closely related to the structure of the
rule table and expressed by the numbers of the rules which break the
strings of the quiescent states.
It is shown that for the N-neighbor and K-state
cellular automata, the class I,
class II, class III and class IV patterns coexist at least in
the range, $\frac{1}{K} \le \lambda \le 1-\frac{1}{K} $.  
The mechanism is studied
quantitatively by introducing a new
parameter $F$, which we call quiescent string dominance
parameter. It is taken to be orthogonal to $\lambda$. 
Using the parameter F and $\lambda$, the
rule tables of one dimensional 5-neighbor and 4-state cellular
automata are classified. 
The distribution of the four 
pattern classes in ($\lambda$,F) plane shows that the rule tables of 
class III pattern class are distributed in larger $F$ region,
while those of class II and class I pattern classes are found in the
smaller $F$ region and the class IV behaviors are observed in the overlap
region between them. These distributions are almost independent of
$\lambda$ at least in the range $0.25 \leq \lambda \leq 0.75$, namely
the overlapping region in $F$, where the class III
and class II patterns coexist, has quite gentle $\lambda$ dependence
in this $\lambda$ region.
Therefore the relation between the pattern classes and the $\lambda$
parameter is not observed.
\end{abstract}

%Use showkeys class option if keyword
\maketitle

\section{Introduction}
Cellular automata (CA) has been one of the most 
studied fields in the research of 
complex systems. 
Various patterns has been generated
by choosing the rule tables.
Wolfram\cite{wolfram} has classified these patterns into four rough
categories:
class I (homogeneous), class II (periodic), class III (chaos) and 
class IV (edge of chaos). 
The class IV patterns have been the most interesting target for the
study of CA, because it provides us with an example of the
self-organization in a simple
system and it is argued that the possibility of 
computation is realized by the complexity at the edge of
chaos\cite{wolframs,langton}.\\
\indent
Much more detailed classifications of CA, have been carried out mainly for
the elementary cellular automata (3-neighbor and 2-state
CA)\cite{hanson,wuensche}, in which the patterns are studied quite
accurately for each rule table. And the classification of the rule
tables are studied by introducing some 
parameters\cite{binder,wuensche,oliveira}. 
However, in this case, there are some confusions in the classification
of the class IV CA. The result on the so to speak
$\rho=1/2$ task from Packard\cite{packard} have been different from
that of Mitchell, Crutchfield and Hraber\cite{mitchell}.
In this case, the number of the independent
rule tables are so small to treat them statistically and the symmetry
of 
the interchange of the states ''0'' and ''1'' make the classifications
of the 
the pattern classes more delicate than those of other CA with $K \geq 3$.
Therefore, 
it may be worth to start with other
models, in order to find the general properties of the  CA.\\
\indent
However the number of rule tables in
N-neighbor and K-state cellular automata CA(N,K) grows like
$K^{K^{N}}$. Therefore
except for a few smallest combinations of the $N$ and $K$, the numbers
of the rule tables become so large that studies of the CA dynamics for
all rule tables are impossible even with the fastest supercomputers. 
Therefore it is important to find a set of parameters by which the pattern
classes could be classified
, even if it is a qualitative one.\\
\indent
Langton has introduced $\lambda$ parameter and
argued that as $\lambda$ increases the pattern class
changes from class I to class II and then to class
III. And class IV behavior is
observed between class II  and class III 
pattern classes\cite{langton0,langton,langton2}.
The $\lambda$ parameter represents rough behavior of CA in the rule table
space, but finally
does not sufficiently classify the quantitative behavior of CA.
%\cite{langton,langton2}.
It is well known that different pattern classes coexist at the same
$\lambda$. 
Which of these pattern classes is chosen, depends on
the random numbers in generating the rule tables.
The reason or mechanism for this is not yet known;  
we have no way to control the pattern classes at fixed $\lambda$.
And the transitions from a periodic to
chaotic pattern classes are observed in a rather wide range of $\lambda$.
In Ref.\cite{langton2}, a schematic phase-diagram
 was sketched.
However a vertical axis was not specified.
Therefore, it has been thought 
that new parameters are necessary to arrive at more quantitative
understandings of the rule table space of the CA
\footnote{In this article, according to the previous authors,
phase diagram and phase transition will be used in
analogy with the statistical physics.}.\\
\indent
In this article, we will clarify the mechanism which discriminate 
the class I, class II, class III and class IV pattern classes at fixed
$\lambda$.
It is closely related to the structure of the rule tables;
%and is related to 
numbers of rules which breaks strings of quiescent states. 
It is studied quantitatively by
introducing a new parameter $F$, which we will call quiescent string
dominance parameter. It is taken to be orthogonal to $\lambda$. 
In the region $1/K \le \lambda \le 1-1/K$, 
the maximum of $F$
corresponds to class III rule tables while minimum of $F$, to class II
or class I rule tables. Therefore
the transition of the pattern classes 
takes place somewhere between these two limits without fail.
By the determination of the 
region of $F$, where the change  of the pattern
classes takes place, we could obtain the phase diagram in
($\lambda$,F) plane, and classify the rule table space.\\
\indent
The determination of the phase diagram is carried out for CA(5,4).
It is found that
the rule tables are not separated by a sharp boundaries but they are
represented by
probability densities. Therefore
we define the equilibrium points of two phases where the  two
probability densities of the pattern classes become equal, and define
the transition region where
the probability densities of the both pattern classes are not too much
different from each other. 
By using the equilibrium points and the transition region, we draw a
phase diagram in ($\lambda$,F) plane.\\
\indent
It is found that $\lambda$ dependences of 
the equilibrium points and transition region are very gentle, and 
they continued to be found at least over the range $0.125 \le \lambda
\le 0.75$. 
It means that all the four pattern classes do coexist over the wide
range in $\lambda$.
Our results for the distributions of these
pattern classes in ($\lambda,F$) plane do not support the well known
relation
between the pattern classes and the $\lambda$ parameter proposed in the
Ref.\cite{langton}.
It will be shown that
the results there, are due to the methods to generate the rule tables
with probability $\lambda$.\\
\indent
In section II, we briefly summarize our notations and present
a key discovery, which leads us to the understanding of the relation
between the structure
of the rule table and pattern classes. It strongly suggested that the
rules which break
strings of the quiescent states play an important role for
the pattern classes.\\
\indent
In section III, the rule tables are classified according to the
destruction and construction of strings of the quiescent states and
we find the method to change the chaotic pattern class into periodic one 
and vice versa while keeping $\lambda$ fixed.
We will show that by using the method, the change of the pattern
classes takes place
without fail, in the region $1/K \le \lambda \leq 1-1/K$.\\
\indent
In section IV, the result obtained in section III is studied 
quantitatively by introducing a new quiescent string dominance
parameter $F$.
It is determined by using the
distribution of class IV rule tables, which we call optimal $F$
parameter.\\
\indent
In section V, using $F$ and $\lambda$, we classify 
the rule tables of CA(5,4) in ($\lambda$,F) plane. It will be shown that
all the four pattern classes do
coexist in  wide range in $\lambda$, contrary to the  result of
Ref. \cite{langton} by Langton. The reason why he obtained his result
will be discussed.\\
\indent
In section VI, the rule tables of CA(5,4) are classified in the
($\lambda$,F) plane, which provides us with the phase diagram.\\
\indent
Section VII is devoted to discussions and conclusions, where a
possibility of the transmission of the initial state information
and the classification of the rule tables by the another intuitive $F$
parameter, will be discussed.\\
\indent 
 
\section{SUMMARY OF CELLULAR AUTOMATA AND A KEY DISCOVERY}
\subsection{Summary of cellular automata }
In order to make our arguments concrete, we focus mainly on 
the one-dimensional 5-neighbor and 4-state CA (CA(5,4)) in the
following, because this is a model in which Langton had argued the
classification of CA by the $\lambda$ parameter.
However the qualitative conclusions in this article, hold true for
general CA(N,K).
These points will be discussed in the subsections III B.\\
\indent
We will briefly summarize our notation of CA\cite{wolfram,langton}.
In our study, the site consists of
150 cells having the periodic boundary condition. The states are denoted
as $s(t,i)$. 
The $t$ represents the time step which takes an integer value,
and the $i$ represents the position of cells which range from $0$ to
$149$. 
The $s(t,i)$ takes values $0,1,2,$ and $3$, and the state $0$ is taken
to be the quiescent state.
The set of the states $s(t,i)$
at a time $t$ is called the configuration.  \\
\indent
The configuration at time $t+1$ 
is determined by that of time $t$ by using following local relation,
\begin{widetext}
\begin{equation}
s(t+1,i) =  T(s(t,i-2),s(t,i-1),s(t,i),s(t,i+1),s(t,i+2)). 
\label{eq:table}
\end{equation}
\end{widetext}
The set of the mappings 
\begin{equation}
T(\mu,\nu,\kappa,\rho,\sigma)=\eta,(\mu,\nu,etc.= 0,1,2,3) 
\label{eq:r_table}
\end{equation}
is called the
rule table.  The rule table consists of $4^5$ mappings,
which is selected from a total of $4^{1024}$ mappings.\\
\indent
The $\lambda$ parameter is defined as\cite{langton0,langton}
\begin{equation}
\lambda=\frac{N_{h}}{1024}, 
\label{eq:lambda}
\end{equation}
where $N_{h}$ is the number in which $\eta$ in Eq.~(\ref{eq:r_table}) is not
equal to $0$. In other words 
the $\lambda$ is the probability that the rules do not select the
quiescent state in next time step. Until section III, we set the
rule tables randomly with the probability $\lambda$. Our method is to
choose $1024-N_{h}$ rules randomly, and set $\eta=0$ in the 
right hand side of Eq.~(\ref{eq:r_table}), and
for the rest of the $N_{h}$ rules, the $\eta$ picks up the number 1,2 and
3 randomly.
The initial configurations are also set randomly.\\
\indent
The time sequence of the configurations 
is called a pattern. The patterns are classified 
roughly into four classes established by Wolfram\cite{wolfram}. 
It is widely accepted that pattern classes are classified by the
$\lambda$\cite{langton};
%has been argued\cite{langton} that 
as the $\lambda$ increases the most frequently generated 
patterns change from homogeneous (class I) to
periodic (class II) and then to chaotic (class III),  
and at the region between class II and class III, the  edge of
chaos (class IV) is located.
\subsection{Correlation between pattern classes and rules 
which break strings of quiescent states} 
\begin{figure*}
\begin{center}
\scalebox{0.60}{ { \includegraphics{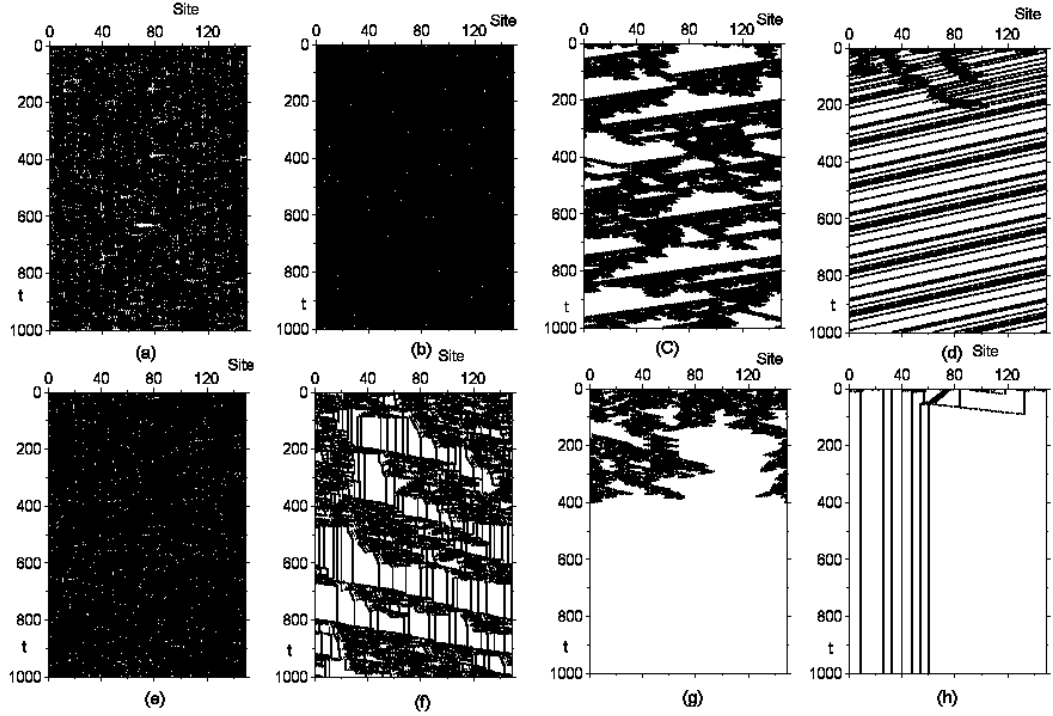} } }
\caption{\label{fig:pattern_at_lambda0.44}
The pattern classes at $N_{h}=450$. The quiescent state is
shown by white dot, while other states are indicated by black
points. Fig.\ref{fig:pattern_at_lambda0.44}(a) corresponds to $N_4=4$,
 Fig.~\ref{fig:pattern_at_lambda0.44}(b),
Fig.~\ref{fig:pattern_at_lambda0.44}(c),
and   Fig.~\ref{fig:pattern_at_lambda0.44}(d), to $N_4=3$, 
Fig.~\ref{fig:pattern_at_lambda0.44}(e),
Fig.~\ref{fig:pattern_at_lambda0.44}(f), and
Fig.~\ref{fig:pattern_at_lambda0.44}(g), to $N_4=2$, and 
Fig.~\ref{fig:pattern_at_lambda0.44}(h) corresponds to $N_4=1$,
respectively.}
\end{center}
\end{figure*}

In order to find the reason why the different pattern classes are
generated at the same $\lambda$, we have started to collect rule
tables of different pattern classes, and tried to
find the differences between them. 
We fix at $N_{h}=450$ ($\lambda=0.44$),
because we find empirically that around this point the chaotic, edge of
chaos, and periodic patterns
are generated with similar ratio.
By changing the random number, we have
gathered a few tens of the rule tables and classified them
into chaotic, edge of chaos, and periodic ones.\\
\indent 
In this article, a pattern is considered the edge of chaos (class IV)
when its transient length\cite{langton} is longer
than $3000$ time steps.\\
\indent
First, we study whether or not the pattern classes are sensitive to the
initial configurations.
We fix the rule tables and change the initial configurations randomly.
For most of the rule tables, 
the details of the patterns depend on the initial configurations, but
the pattern classes are not changed\cite{wolfram}. The exceptions
will be discussed in the subsection VII A, in connection with the
transmission of the initial state information.
Thus the differences of the pattern classes are due to the differences 
in the rule tables, and the target of our inquiry has to do with the
differences between them.\\
\indent
For a little while, we do not impose a quiescent
condition (QC),
%\cite{langton},
$T(0,0,0,0,0)=0$,
because without this condition, 
the structure of the rule table becomes more transparent.
This point will further be discussed in  section IV.\\
\indent
After some trial and error, we have found a strong correlation between
the pattern classes and the QC.
The probability of the rule table,
which satisfies the QC is much larger in class II patterns than that
in the class III
patterns. This correlation suggest that the rule 
$T(0,0,0,0,0)=h,\hspace{0.2cm} h \neq 0$, which breaks the string of the 
quiescent states with length 5, 
pushes the pattern toward chaos.
We anticipate that the similar situation will hold for the length 4
strings of quiescent states.\\
\indent
We go back to the usual definitions of CA; in the following we discuss 
CA under QC, $T(0,0,0,0,0)=0$. 
We study the correlation between the number of the rules which breaks
the length four strings of the quiescent states. These rules are given
by,
\begin{equation}
\begin{array}{ll}
T(0,0,0,0,i)=h,\\
T(i,0,0,0,0)=h,(i,h=1,2,3).
\label{eq:d4}
\end{array}
\end{equation}
They will also push the pattern toward chaos.
%\footnote{
Similar properties of the rule tables had
been noticed by Wolfram and Suzudo with the
arguments of the unbounded growth\cite{wolfram} and
expandability\cite{suzudo}.\\
%}.
\indent
We denote the total number of rules of Eq.~(\ref{eq:d4}) in a rule table 
as $N_4$. In order to study the correlation between the pattern
classes and the number $N_4$,
we have collected 30 rule tables and grouped them by the number $N_4$.
We have 4 rule tables with $N_4 \ge 4$ , 13 rule tables with $N_4=3$, 9
rule tables with $N_4=2$ and 4 rule tables with $N_4 \le 1$.
When $ N_4 \ge 4$, all rule tables generate 
chaotic patterns, while when $N_4 \le 1$, only
periodic ones are generated.
At $N_4=3$ and $N_4=2$, chaotic, edge of chaos, and periodic
patterns coexist.
Examples are shown in the Fig.~\ref{fig:pattern_at_lambda0.44}.
The coexistence of three pattern classes at $N_4=3$ is seen in
Fig.~\ref{fig:pattern_at_lambda0.44}(b),
Fig.~\ref{fig:pattern_at_lambda0.44}(c) and Fig.~\ref{fig:pattern_at_lambda0.44}(d)
and that of $N_4=2$ is exhibited in 
Fig.~\ref{fig:pattern_at_lambda0.44}(e),
Fig.~\ref{fig:pattern_at_lambda0.44}(f) and Fig.~\ref{fig:pattern_at_lambda0.44}(g).\\
\indent
As anticipated, the strong correlation between $N_4$
%, the number of rules of
%Eq.~\ref{eq:d4} in a rule table,
 and the pattern classes has been
observed in this case too.
These discoveries have provided us with a key hint leading us to the
hypothesis 
that the rules, which break strings of the quiescent states, will
play a major role for the pattern classes.

\section{Structure of rule table and pattern classes}
\subsection{Structure of rule table and replacement experiment}
\begin{figure*}
\begin{center}
\scalebox{0.60}{ { \includegraphics{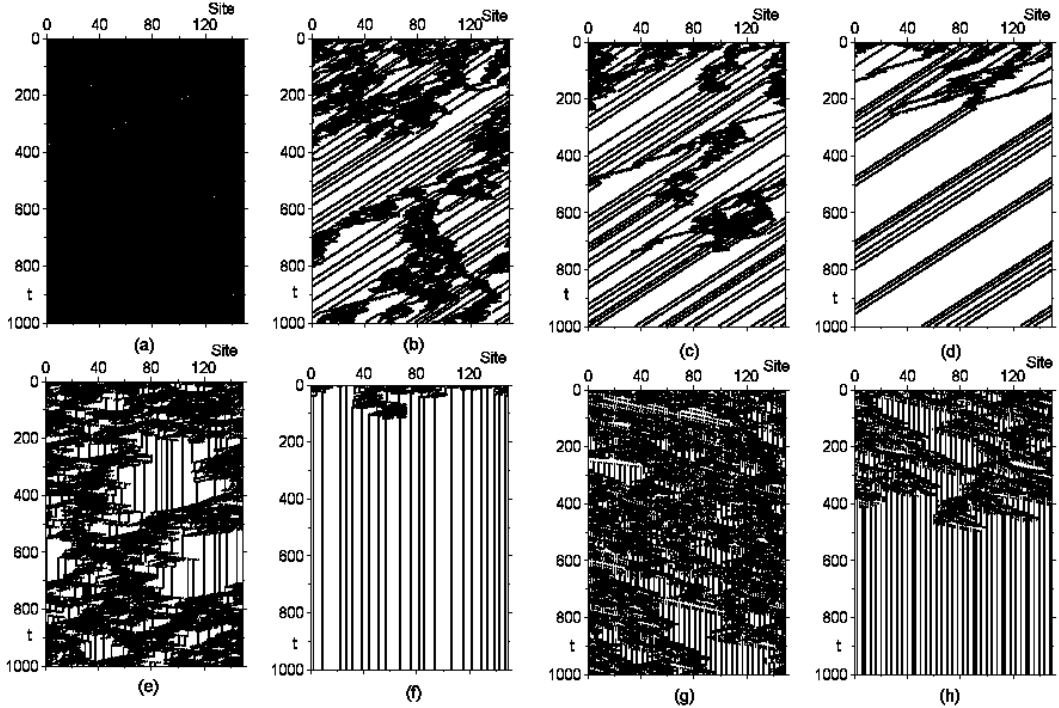} } }
\caption{
An example of the replacement experiments at $N_h=615$.
Fig.~\ref{fig:replace1}(a) is obtained randomly with $N_h=615$
($\lambda=0.6)$, by the method explained in subsection II A. 
Fig.~\ref{fig:replace1}(b) to Fig.~\ref{fig:replace1}(h) are obtained by the
replacements of the rule table
of Fig.~\ref{fig:replace1}(a), which are summarized in Table~\ref{tab:table2}.
}
\label{fig:replace1}
\end{center}
\end{figure*}

In order to test the hypothesis of the previous section,
we classify the rules into four groups according to the
operation on strings of the quiescent states.
In the following, Greek characters in the rules
represent groups $0,1,2,3$ while Roman, represent groups $1,2,3$.\\
\noindent
Group 1: $T(\mu,\nu,0,\rho,\sigma)=h$.\\
The rules in this group break strings of the quiescent states.\\
Group 2: $T(\mu,\nu,0,\rho,\sigma)=0$. \\
The rules of this group conserve them.\\
Group 3: $T(\mu,\nu,i,\rho,\sigma)=0$.\\
The rules of this group develop them.\\
Group 4: $T(\mu,\nu,i,\rho,\sigma)=l$. \\
The rules in this group do not affect string of quiescent states in next 
time step.\\
\indent
Let us denote the number of the group 1 rules in a rule table as
$N(g1)$. Similarly for the number of other groups. They satisfy the
following sum rules, when $N_h$ is fixed.
\begin{equation}
\begin{array}{ll}
N(g1)+N(g2)=256,\\
N(g3)+N(g4)=768,\\
N(g2)+N(g3)=1024-N_h,\\
N(g1)+N(g4)=N_h.
\end{array}
\label{eq:group}
\end{equation}
In the methods of generating the rule tables randomly using $N_h$
($\lambda$), these numbers are determined mainly by the probability
$\lambda$; namely $N(g1) \simeq 256 \lambda$, $N(g2) \simeq 256(1-\lambda)$
$N(g3) \simeq 768(1-\lambda)$ and $N(g4) \simeq 768 \lambda$
respectively.
Therefore they suffer from fluctuation due to randomness.\\
\indent
The group 1 rules are further classified into five types according to
the length of string of quiescent states, which they break. These are
shown in Table~\ref{tab:destruc}. 
The rule D5 is always excluded from rule tables by the 
quiescent condition.\\
\renewcommand{\arraystretch}{0.9}
\begin{table*}[h]
\caption{ The classification of the rules in group 1 into five types,
where $h \neq 0$.}
%\vspace*{0.3cm}
\label{tab:destruc}
\begin{center}
\begin{tabular}{|c|c|c|c|c|c|c|}
     \hline
     \multicolumn{1}{|c|}{type} &
     \multicolumn{1}{|c|}{Total Number} &
     \multicolumn{1}{|c|}{Name}&
     \multicolumn{1}{|c|}{Replacement}\\
     \hline
        $T(0,0,0,0,0)=h$  &1  &D5  &RP5,RC5\\
     \hline
        $T(0,0,0,0,i)=h$  &3  &D4  &RP4,RC4\\
        $T(i,0,0,0,0)=h$  &3  &    &    \\
     \hline
        $T(0,0,0,i,\sigma)=h$ &12 &   &    \\
        $T(i,0,0,0,m)=h$      &9  &D3 &RP3,RC3\\
        $T(\mu,j,0,0,0)=h$    &12 &   &    \\
     \hline
        $T(\mu,j,0,0,m)=h$    &36 &D2 &RP2,RC2 \\
        $T(i,0,0,l,\sigma)=h$ &36 &   &       \\
     \hline
        $T(\mu,j,0,l,\sigma)=h$ &144 &D1 &RP1,RC1\\
     \hline
\end{tabular}  
\end{center}  
\end{table*}
\indent
Our hypothesis presented at the end of the
section II is expressed more quantitatively as follows;
the numbers of the D4, D3, D2, and D1 rules shown in Table~\ref{tab:destruc} 
will mainly determine the pattern classes.\\
\indent
In order to test this hypothesis, we artificially change the numbers of
the rules in Table~\ref{tab:destruc} while keeping
the $N_h$ ($\lambda$) fixed. 
For D4 rules, we carry out the replacements defined by the following
equations, 
\begin{equation}
\begin{array}{ll}
T(0,0,0,0,i)=h \rightarrow T(0,0,0,0,i)=0, \\ 
or \hspace {0.1cm}T(i,0,0,0,0)=h \rightarrow T(i,0,0,0,0)=0, \\ 
T(\mu,\nu,j,\rho,\sigma)=0 \rightarrow T(\mu,\nu,j,\rho,\sigma)=l,\\
\label{eq:toperio} 
\end{array}
\end{equation}
where except for $h$, the groups $\mu$, $\nu$, $\rho$, $\sigma$, $j$ and $l$ 
are selected randomly.
Similarly the replacements are generalized for D3, D2, and D1 rules
, which are denoted as RP4 to RP1 in Table~\ref{tab:destruc}. 
They change the rules of group 1 to
that of group 2 together with group 3 to group 4 and are expected
to push the rule table toward the periodic direction. \\
\indent
The reverse replacements for D4 are
\begin{equation}
\begin{array}{ll}
T(0,0,0,0,i)=0 \rightarrow T(0,0,0,0,i)=h, \\
or \hspace {0.1cm}T(i,0,0,0,0)=0 \rightarrow T(i,0,0,0,0)=h, \\
T(\mu,\nu,j,\rho,\sigma)=l \rightarrow T(\mu,\nu,j,\rho,\sigma)=0,\\
\label{eq:tochaos} 
\end{array}
\end{equation} 
which will push the rule table toward the chaotic direction. In this case,
the groups $h$, $\mu$, $\nu$, $j$, $\rho$, and $\sigma$ are
selected randomly.
Similarly we introduce the replacements for D3, D2, and D1, which will
be called RC4 to RC1 in the following.\\
\indent
By the replacement of RP4 to RP1 or 
RC4 to RC1, 
we change the numbers of the rules in Table
\ref{tab:destruc} while keeping the $N_h$ fixed. We denote these
numbers $N_4$, $N_3$, $N_2$, and $N_1$ 
for D4, D3, D2, and D1 rules, respectively. By applying these
replacements, we could study the rule tables which are difficult to
obtain using only $N_h$ ($\lambda$).\\
\indent 
Examples of the replacement experiments are shown in
Fig.~\ref{fig:replace1}.
\renewcommand{\arraystretch}{0.9}
\begin{table*}[h]
\begin{center}
\caption{ The numbers of the group 1 rules and numbers of the
replacements for the rule table of Fig.~\ref{fig:replace1}(a) to
change the patten classes shown in Fig.~\ref{fig:replace1}}
\label{tab:table2} 
%\vspace*{0.3cm}
\begin{tabular}{|c|c|c|c|c|c|c|c|c|c|c|c|c|}
     \hline
     \multicolumn{1}{|c|}{Figure} &
     \multicolumn{1}{|c|}{$N_4$}&
     \multicolumn{1}{|c|}{$N_3$}&
     \multicolumn{1}{|c|}{$N_2$}&
     \multicolumn{1}{|c|}{$N_1$}&
     \multicolumn{1}{|c|}{RP4}&
     \multicolumn{1}{|c|}{RP3}&
     \multicolumn{1}{|c|}{RP2}&
     \multicolumn{1}{|c|}{RP1}\\
     \hline
        Fig.~\ref{fig:replace1}(a)  &3 &22 &53 &96 &0 &0 &0 &0\\
     \hline
        Fig.~\ref{fig:replace1}(b)  &0 &14 &53 &96 &3 &8 &0 &0\\
     \hline
        Fig.~\ref{fig:replace1}(c)  &0 &13 &53 &96 &3 &9 &0 &0\\
     \hline
        Fig.~\ref{fig:replace1}(d)  &0 &12 &53 &96 &3 &10 &0 &0\\
     \hline
        Fig.~\ref{fig:replace1}(e)  &1 &7 &53 &96 &2 &15 &0 &0\\
     \hline
        Fig.~\ref{fig:replace1}(f)  &1 &6 &53 &96 &2 &16 &0 &0\\
     \hline
        Fig.~\ref{fig:replace1}(g)  &2 &1 &53 &96 &1 &21 &0 &0\\
     \hline
        Fig.~\ref{fig:replace1}(h)  &2 &0 &53 &96 &1 &22 &0 &0\\
     \hline
\end{tabular}  
\end{center}  
%\vspace{0.5cm}
\end{table*}
In this replacements, the RP4s are always
carried out first, after that RP3s are done.
The rule table of Fig.~\ref{fig:replace1}(a) is obtained randomly with
$N_h=615$ ($\lambda=0.6)$. 
The numbers of the D4, D3, D2 and D1 are
shown in line  Fig.~\ref{fig:replace1}(a) of Table~\ref{tab:table2}. 
At $N_h=615$ most of the randomly obtained rule tables
generate chaotic patterns.
We start to make RP4 tree times, then number of the D4 becomes
$N_4=0$. At this stage, the rule table still generate chaotic
patterns. Then we proceed to carry out RP3. The chaotic patterns
continues from $N_3=22$ to $N_3=15$, and when $N_3$ becomes $14$, the
pattern changes to edge of chaos behavior, which is shown in the 
Fig.~\ref{fig:replace1}(b).
Fig.~\ref{fig:replace1}(c) is obtained by one more RP3 replacements for
Fig.~\ref{fig:replace1}(b) rule table. It shows a
periodic pattern with a rather long transient length. The pattern with
one more
replacement of RP3 for the Fig.~\ref{fig:replace1}(c), is shown in
Fig.~\ref{fig:replace1}(d),  where the transient length
becomes shorter.
These numbers of D4 rule ($N_4$) and D3 rule ($N_3$), and numbers of the
replacements RP4s and  RP3s for the Fig.~\ref{fig:replace1}(a) rule
table, are summarized in Table~\ref{tab:table2}.\\
\indent 
Similarly replacement experiments in which the RP4s are stopped at
$N_4=1$ and $N_4=2$ are shown
in Fig.~\ref{fig:replace1}(e), Fig.~\ref{fig:replace1}(f), and
Fig.~\ref{fig:replace1}(g), Fig.~\ref{fig:replace1}(h), respectively
and the numbers of group 1 rules and the replacements are also
summarized in the
corresponding lines in the Table~\ref{tab:table2}.
We should like to notice that the transitions to class III to class IV
pattern classes take place at ($N_4=0$, $N_3=13$), ($N_4=1$,$N_3=6$)
and ($N_4=2$,$N_3=0$). Therefore the effects to push the rule table
toward chaos is stronger for D4 than D3 of
Table~\ref{tab:destruc}.
This point will be discussed more
quantitatively in the next section.\\
\indent  
At $N_h=$819, $768$, $717$, $615$, $512$, $410$, $307$ 
and $205$ ($\lambda=0.8$, 0.75, 0.7, 0.6, 0.5, 0.4, 0.3 and 0.2), we
have  carried out replacements experiments for 119, 90,
90, 117, 107, 92, 103 and 89 rule tables, which are generated
randomly using $N_h$ ($\lambda$).
At all the $N_h$ points, we have succeeded in changing the 
the patten classes from class III to class II or class I or vice
versa, by changing the numbers of the group 1 rules.
And in many cases, the edge of chaos behaviors are observed between
them.

\subsection{Chaotic and periodic limit of general CA(N,K)}
Let us study the effects of the replacements theoretically in general
cellular automata CA(N,K).
In the general case too, the rule tables are classified into
four groups as shown in subsection III A.
We have denoted these numbers as $N(g1)$, $N(g2)$, $N(g3)$ and
$N(g4)$.
When $N_h$ is fixed, these numbers satisfy the following sum rules,
which are the generalization of Eq.~\ref{eq:group}.
\begin{equation}
\begin{array}{ll}
N(g1)+N(g2)=K^{N-1},\\
N(g3)+N(g4)=K^{N-1}(K-1),\\
N(g2)+N(g3)=K^{N}-N_h,\\
N(g1)+N(g4)=N_h.
\label{eq:sum_rule1}
\end{array}
\end{equation}
\indent
However the individual number $N(gi)$ suffers from the fluctuations
due to randomness.
%If rule tables are generated randomly only by using $\lambda$ ($N_h$),
They distribute with the mean given by Table~\ref{tab:N-K}.
In this subsection, let us neglect these fluctuations.\\
\renewcommand{\arraystretch}{0.9}
\begin{table*}[h]
\caption{The classification of CA(N,K) rules into 4 groups.
The $\mu_{i}$ represent 0 to K-1 while $h$,$i$ and $l$,
1 to K-1.}
%\vspace*{0.3cm}
\label{tab:N-K}
\begin{center}
\begin{tabular}{|c|c|c|c|c|c|c|}
     \hline
     \multicolumn{1}{|c|}{ } &
     \multicolumn{1}{|c|}{rule } &
     \multicolumn{1}{|c|}{Average number}\\
     \hline
        group 1  &$T(\mu_1,\mu_2,...,0,...,\mu_{N})=h$  &$K^{N-1}\lambda$\\
     \hline
        group 2  &$T(\mu_1,\mu_2,...,0,...,\mu_{N})=0$  &$K^{N-1}(1-\lambda)$\\
     \hline
        group 3  &$T(\mu_1,\mu_2,...,i,...,\mu_{N})=0$  &$K^{N-1}(K-1)(1-\lambda)$\\
     \hline
        group 4  &$T(\mu_1,\mu_2,...,i,...,\mu_{N})=l$  &$K^{N-1}(K-1)\lambda$\\
     \hline
\end{tabular}  
\end{center}  
\end{table*}
\indent
The replacements to decrease the number of the group 1 rule 
while keeping the $\lambda$ fixed are given by,
\begin{equation}
\begin{array}{ll}
N(g1) \rightarrow N(g1)-1,\hspace {0.5cm} N(g2) \rightarrow N(g2)+1,\\
N(g3) \rightarrow N(g3)-1,\hspace {0.5cm} N(g4) \rightarrow N(g4)+1.\\
\label{eq:del_num_RP}
\end{array} 
\end{equation}
In CA(5,4), they correspond to RP4 to RP1 of section III A.\\
\indent
These replacements stop either when $N(g1)=0$ or $N(g3)=0$ is reached.
Therefore when $N(g1) \le N(g3)$, which corresponds to $\lambda \le
(1-\frac{1}{K})$ in $\lambda$,
all the group 1 rules are replaced by the group 2 rules. 
In this limit, quiescent
states at time $t$ will never be changed, because there is no rule
which 
converts them to other states, while the group 3 rules have a chance to
create a new quiescent state in the next time step. Therefore
the number of quiescent states at time t is a non-decreasing function
of t. Then, the pattern class should be class I (homogeneous) or 
class II (periodic), which we call periodic limit. Therefore
the replacements of Eq.~(\ref{eq:del_num_RP}) push the rule table toward the
periodic direction.\\ 
\indent
 Let us discuss the reverse replacements of Eq.~(\ref{eq:del_num_RP}).
In these replacements, if $N(g2) \le N(g4)$, which corresponds to 
$\frac{1}{K} \le \lambda$,
all group 2 rules are replaced by
the group 1 rules, except for the quiescent condition. 
In this extreme reverse case, 
almost all the quiescent states at time t are
converted to other states in next time step, while group 3 rules will
create them at different places. 
Then this will most probably develop into chaotic patterns. This limit
will be called chaotic limit.\\
\indent
We should like to say that there are possibilities that 
atypical rule tables and  initial conditions might
generate a periodic patterns even in this limit. But in this article, these
atypical cases are not discussed.\\
\indent
Therefore in the region,
\begin{equation} 
\frac{1}{K} \le \lambda \le 1-\frac{1}{K},
\label{eq:2-limit}
\end{equation}
all the rule tables are located somewhere between
these two limit, and by the successive
replacements of Eq.~(\ref{eq:del_num_RP}) and their reverse ones, the changes 
of the pattern classes take place without fail.
This is the theoretical foundation of the replacement experiments 
of previous subsection and also explains why in this region the four
pattern classes coexist.\\
\indent
The Eq.~(\ref{eq:del_num_RP}) provide us with a method to control the
pattern classes at fixed $\lambda$. 
The details of the replacements of Eq.~(\ref{eq:del_num_RP})
depend on the models. In the CA(5,4), they have been 
RP4 to RP1 and RC4 to RC1. They will enable us to obtain a rule tables
which are difficult to generate by the method ''random-table
method'' or ''random-walk-through method'' and lead us to the new
understandings on the structure of 
CA rule tables in section V.

\section{Quiescent String Dominance Parameter $F$ in CA(5,4)}
In the previous section, we have found that each rule table is located
somewhere between chaotic limit and periodic limit,
in the region $\frac{1}{K} \le \lambda \le 1-\frac{1}{K}$. 
In order to express
the position of the rule table quantitatively, we introduce a new
quiescent string dominance 
parameter F, which provides us with a new axis ($F$-axis) orthogonal to
$\lambda$.  Minimum of $F$ is the periodic limit, while maximum of it
corresponds to chaotic limit. In this section, we will
determine the parameter $F$ for CA(5,4).\\
\indent
As a first approximation, the parameter $F$ is taken to be 
be a function of
the numbers of the rules D4, D3, D2 and D1, which have been denoted as
$N_4$, $N_3$, $N_2$ and $N_1$, respectively.  
We proceed to
determine $F(N_4,N_3,N_2,N_1)$ 
by applying simplest approximations and assumptions\\
\indent
 We apply Taylor series expansion for $F$, and
approximate it by the linear terms in $N_4$, $N_3$ $N_2$ and $N_1$. \\
\begin{equation}
 F(N_4,N_3,N_2) \simeq c_4 N_4 + c_3 N_3 + c_2 N_2 + c_1 N_1
 \label{eq:taylor}.
\end{equation} 
where $c_4= \partial F/\partial N_4$, similar for $c_3$, $c_2$ and $c_1$.
They represent the strength of the effects of the rules
D4, D3, D2 and D1 to push the rule table toward chaotic direction.
These definitions are symbolic, because $N_{i}$ is discrete.\\
\indent
 The measure in the $F$ is still arbitrary. We fix it in the unit
where the increase in one unit of $N_4$ results in the  change of $F$ in one
unit. This corresponds to divide $F$ in Eq.~(\ref{eq:taylor}) by $c_4$, and to
express it by the ratio $c_3/c_4$ ($r_3$), $c_2/c_4$ ($r_2$) and
$c_1/c_4$ ($r_1$).\\
\indent
Before we proceed to determine $r_3$, $r_2$ and $r_1$, let us interpret the
parameter $F$ geometrically. Most generally, the rule tables are
classified in 1024-dimensional space in CA(5,4). The location of the
rule table of each pattern classes forms a hyper-domain in this space. 
We map the points in the  hyper-domain
into 4-dimensional $(N_4,N_3,N_2,N_1)$ space, in which
they will also be located in some region.
%in the 4-dimensional space. 
We introduce a surface $S(N_4,N_3,N_2,N_1)=\Phi$
in order to line up these points. $F$-axis is a normal line
of this surface.
In Eq.~(\ref{eq:taylor}), we approximate it by a hyper plane. \\
\indent
In order to determine $r_3$, $r_2$ and $r_1$,
we apply an argument that the class IV rule tables are located around the
boundary of the class II and class III rule tables. 
Our strategy to determine $r_3$, $r_2$ and $r_1$
is to find the regression hyper plane of class IV rule tables on four
dimensional space, ($N_4$,$N_3$, $N_2$ $N_1$). 
It is equivalent to fix the $F$-axis in such a way that
the projection of the distribution of class IV rule tables on $F$-axis,
looks as narrow as possible. 
The quality of our approximations and
assumptions reflects the width of the distribution of class IV rule
tables.\\
\indent
In the least square method, our problem is formulated to find $r_3$,
$r_2$ and $r_1$, which minimize the quantity,
\begin{widetext}
\begin{equation}
s(r_3,r_2,r_2)= 
\frac{1}{c_{4}^{2}}\sum_{i,j}(F_{class IV}^i(N_4^i,N_3^i,N_2^i,N_1^i)
            - F_{class IV}^j(N_4^j,N_3^j,N_2^j,N_1^j))^{2}. 
\label{eq:ansatz},
\end{equation}
\end{widetext}
where $i$ and $j$ label the class IV rule tables.
We solve the equations,
$\partial S/\partial r_3=0$, $\partial S/\partial r_2=0$ and $\partial
S/\partial r_1=0$,
which are
\begin{widetext}
\begin{equation}
\begin{array}{llll}
\displaystyle
 r_3 \sum_{i,j} (\delta N_3^{i,j})^2 +
    r_2 \sum_{i,j} \delta N_2^{i,j}\delta N_3^{i,j} +
      r_1 \sum_{i,j} \delta N_1^{i,j}\delta N_3^{i,j}
         =- \sum_{i,j} \delta N_4^{i,j}\delta N_3^{i,j}, \\
\displaystyle
 r_3 \sum_{i,j} \delta N_3^{i,j} \delta N_2^{i,j} +
    r_2 \sum_{i,j} (\delta N_2^{i,j})^2+
      r_1 \sum_{i,j} \delta N_1^{i,j} \delta N_2^{i,j}
         =- \sum_{i,j} \delta N_4^{i,j}\delta N_2^{i,j}, \\

\displaystyle
 r_3 \sum_{i,j} \delta N_3^{i,j} \delta N_1^{i,j} +
    r_2 \sum_{i,j} \delta N_2^{i,j} \delta N_1^{i,j}+
      r_1 \sum_{i,j} (\delta N_1^{i,j})^2
         =- \sum_{i,j} \delta N_4^{i,j}\delta N_1^{i,j}. \\
\label{eq:sol_r3}
\end{array}
\end{equation}
\end{widetext}
where $\delta N_4^{i,j}=N_{4}^{i}-N_{4}^{j}$, similar for
$\delta N_3^{i,j}$ and $\delta N_2^{i,j}$.\\
\indent
In order to collect class IV rule tables, we have generated rule
tables randomly both for $N_h$ in the region, 
$205 \le N_h \le 819$ ($0.2 \le \lambda \le 0.8$) and for the numbers
of the group 1 rules in the ranges,
$0 \le N_4 \le 6$, $0 \le N_3 \le 33$, $0 \le N_2 \le 72$ and
$0 \le N_1 \le 144$.\\
\indent
This is realized by the two step method.
 In the first step, we generate rule table randomly using the  number
$N_h$, which are explained in subsection II B.  
We should like to notice that under this methods, the numbers of the
group 1 rules,
 $N_4$, $N_3$, $N_2$ and $N_1$
are distributed around, 
$6 \lambda$, $33 \lambda$, $72 \lambda$ and $144 \lambda$ respectively.
They are denoted as $N_i^{\lambda}$.
Then in second step, $N_i$s are determined randomly
between zero and their maximum.
The $N_i^{\lambda}$'s, which are obtained in the first step are
changed to their random value
by the replacements RC4 to RC1 or RP4 to RP1.\\
\indent
We have generated about 14000 rule tables, and classify them into four
pattern classes according to their transient length. There are
483 class I, 3169 class II, 10248 class III and 329 class IV 
rule tables, respectively.
From the 329 Class IV rule tables, the coefficients $r_i$
are determined by solving the Eq.~(\ref{eq:sol_r3}). 
They are summarized in the Table~\ref{tab:coefficient-r}, where the errors
are estimated by the Jackknife method.
\renewcommand{\arraystretch}{0.9}
\begin{table}[h]
\caption{The optimal and intuitive coefficients $r_i$.}
%\vspace*{0.3cm}
\label{tab:coefficient-r}
\begin{center}
\begin{tabular}{|c|c|c||c|c|c|c|}
     \hline
     \multicolumn{1}{|c|}{ } &
     \multicolumn{1}{|c|}{optimal}&
     \multicolumn{1}{|c||}{Error}&
     \multicolumn{1}{|c|}{Intuitive}\\
     \hline
         $r_3$  &0.1563 &0.0013 &0.18182\\
     \hline
         $r_2$  &0.0506 &0.0007 &0.08333\\
     \hline
         $r_1$  &0.0195 &0.0002 &0.04167\\
     \hline
\end{tabular}  
\end{center}  
\end{table}
The results show that the coefficients are positive, and satisfy the
order,
\begin{equation}
 c_4 > c_3 > c_2 > c_1.  
\label{eq:order_force}
\end{equation} 
It means that the effects to move the rule table toward chaotic limit
on the
$F$-axis are stronger for the rules which break longer strings of the
quiescent states
\footnote{
If the quiescent
condition is not imposed,
 Eq.~(\ref{eq:order_force}) will becomes 
$$
%\begin{equation}
c_5 > c_4 > c_3 > c_2 > c_1. 
%\end{equation}
$$
Therefore the correlation between 
pattern classes and the existence of D5 rule is stronger than that
between those and the number of D4 rules.
If we start our 
study within the quiescent condition, we may make a longer
detour to
find the hypothesis of section II and get the qualitative conclusion of
section III.}.\\
\indent
The order in Eq.~(\ref{eq:order_force}) is understood
by the following intuitive arguments.
If six D4 rules are included in the rule table, the string of the 
quiescent states with length 5 will not develop. Similarly, 
if 33 D3 rules are present in the rule table, there is no chance
that the length 
4 string of the quiescent states could be made. These are roughly 
similar situations for the formation of pattern classes.
Thus the strength of the D3 rules $r_3$ will be roughly
equal to $6/33$ of that
of D4 rules, similar for the strength of the D2 and D1 rules. These
coefficients are
also shown in the Table~\ref{tab:coefficient-r}.
We call the $F$ parameter with these
$r_{i}$'s as intuitive $F$ parameter and those determined
by solving the Eq.~(\ref{eq:sol_r3}) as optimal one. 
It is found that the differences between them are not so large. 
The classification of
the rule tables with intuitive $F$ parameter will be discussed in the
subsection VII B.\\
\section{Distribution of the Rule Tables in ($\lambda$,F) Plane}
\begin{figure*}
\begin{center}
\scalebox{0.85}{ { \includegraphics{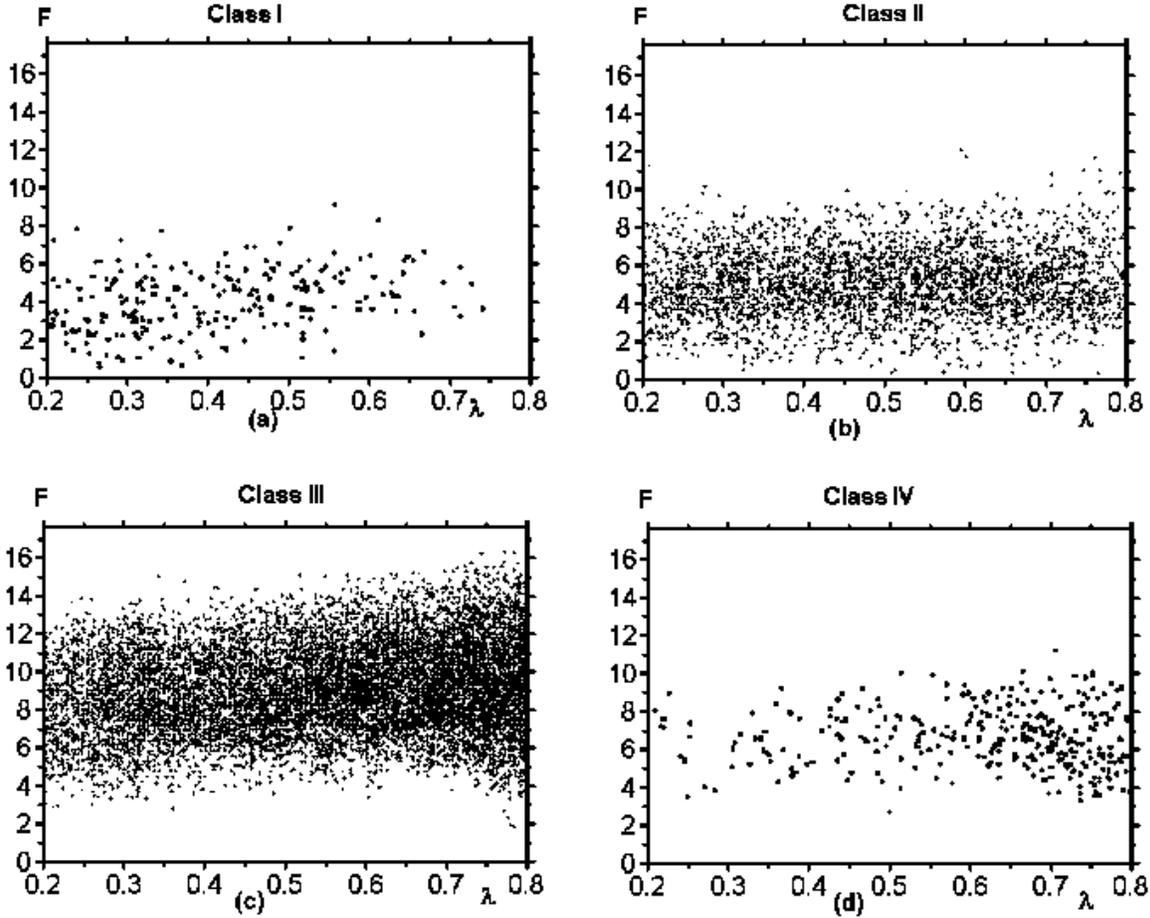} } }
\caption{\label{fig:lambda-F-distribution} 
Distribution of rule tables in ($\lambda$,F) plane.
Fig.~\ref{fig:lambda-F-distribution}(a): The distribution of 483 class I
rule tables.
Fig.~\ref{fig:lambda-F-distribution}(b): The distribution of 3169 class II
rule tables.
Fig.~\ref{fig:lambda-F-distribution}(c): The distribution of 10248 class
III rule tables.
Fig.~\ref{fig:lambda-F-distribution}(e): The distribution of 329 class
IV rule tables.
}
\end{center}
\end{figure*}

Using the optimal $F$ parameter we plot the position of the rule tables of
each pattern classes in ($\lambda$,F) plane. They are shown in
Fig.~\ref{fig:lambda-F-distribution}.\\
\indent
The Figs.~\ref{fig:lambda-F-distribution} shows that
the class III rule tables are located
in the larger $F$ region; about $F \geq 4$, while class I, class II
rule tables,
in the smaller $F$ region; about $F \leq 9$, and the class
IV rule tables are found in the overlap region of class II and
class III rule tables; about $4 \leq F \leq 9$. 
These results support the chaotic limit and periodic limit discussed
in the subsection III B, and shows that at least   
in the $0.2 \leq \lambda \leq 0.75$ range,
all four pattern classes coexist.\\
\indent 
These distributions of rule table in $F$, are almost
independent of $\lambda$.
It means that 
the CA pattern classes are
not classified by $\lambda$, 
contrary to the results of Ref. \cite{langton}, but are 
classified rather well by the quiescent string dominance parameter
$F$.\\
\indent
Let us discuss the reason why Langton had obtained his results.  
If the rule tables are generated using the probability $\lambda$ by the 
''random-table method'' or ''random-walk-through
method''\cite{langton}, the numbers of the group 1 rule tables $N_4$,
$N_3$, $N_2$ and $N_1$ are also controlled by the probability $\lambda$.
They distribute around $N^{\lambda}_{i}$ of section IV.
Then the $F$ parameters are also distributed around,
\begin{equation}
F^{\lambda} = (6+33r_3+72r_2+144r_1)\lambda.
\label{eq:F_with_lambda}
\end{equation}
\indent
Therefore in these methods, $\lambda$ and  $F$ are strongly
correlated.
The probabilities to obtain the rule tables, which are far apart
from the line given by Eq.~(\ref{eq:F_with_lambda}) are very small.
When $\lambda$ is small, the rule table with small $F$ are mainly
generated, which are class I and class II CAs. On the other
hand in the large $\lambda$ region, the rule tables with large
$\lambda$ are dominantly generated, which are class III CAs.
The line of Eq.~(\ref{eq:F_with_lambda}) crosses the location of class
IV pattern classes around $0.4 \leq \lambda \leq 0.55$.
This might be a reason Langton has obtained his results.
But the distribution of rule tables in all ($\lambda$,F) plane, show
the global structure of the CA
rule table space as in Fig.~\ref{fig:lambda-F-distribution} and
lead us to
deeper understanding of the structure of CA rule tables.\\ 
\indent
We should like to stress again that four pattern classes do coexist in a
rather wide range in $\lambda$, which are rather well classified by
the parameter $F$ not by $\lambda$.

\section{Classification of Rule tables in ($\lambda$,F) Plane}
In the Fig.~\ref{fig:lambda-F-distribution}, it is found that rule tables
are not separated by sharp boundaries. And they seem to have some
probability distributions. 
We denote the probability densities of class I, class II, class III and
class IV pattern classes as $P^{I}$, $P^{II}$, $P^{III}$ and $P^{IV}$,
respectively and proceed to
classify the rule tables by using them.\\
\indent
The equilibrium points $F^{II-III}_{E}(\lambda)$ of class II and
class III rule tables are defined by the point where the relation
$P^{II}(F)=P^{III}(F)$ is satisfied.
The region in ($\lambda$,$F$) plane where $P^{II}$ and $P^{III}$ 
coexist in a similar ratio is defined as transition region.
The upper points of the transition region $F^{II-III}_{U}$, are defined
by the points, $P^{II}$=$\frac{P^{III}}{e}$ 
and similarly for 
the lower points of the transition region $F^{II-III}_{L}$, where
$P^{II}$ and $P^{III}$ are interchanged. By these three points,
$F^{II-III}_{E}$, $F^{II-III}_{U}$ and $F^{II-III}_{L}$, we define the
phase boundary of the rule tables.\\
\indent
The distributions of the rule tables in the
Fig.~\ref{fig:lambda-F-distribution} show the 
qualitative probability distributions. However in order to study the
$\lambda$ dependences of $F^{II-III}_{E}$, $F^{II-III}_{U}$ and
$F^{II-III}_{L}$ more quantitatively, we 
generate a rule tables at fixed $\lambda$s.
The $\lambda$ points and numbers of
the rule tables are shown in Table~\ref{tab:phtr_data}.
\renewcommand{\arraystretch}{0.9}
\begin{table*}[h]
\caption{\label{tab:phtr_data}
 The data points, numbers of each pattern classes and 
the transition parameters of CA(5,4). The column ''Comp'' shows the
numbers of rule tables
which transmit the information of initial states.
 }
%\vspace*{0.3cm}
\begin{center}
\begin{tabular}{|c|c|c|c|c|c|c|c|c|c|c|c|c|}
 \hline
     \multicolumn{2}{|c|}{ }&
     \multicolumn{5}{|c|}{Number of the rule tables}&
     \multicolumn{3}{|c|}{Class I - II }&
     \multicolumn{3}{|c|}{Class II - III }\\
     \hline
     \multicolumn{1}{|c|}{$\lambda$} &
     \multicolumn{1}{|c|}{$N_{h}$} &
     \multicolumn{1}{|c|}{I} &
     \multicolumn{1}{|c}{II}&
     \multicolumn{1}{|c|}{III}&
     \multicolumn{1}{|c|}{IV}&
     \multicolumn{1}{|c|}{Comp}&
     \multicolumn{1}{|c|}{$F^{I-II}_{L}$}&
     \multicolumn{1}{|c|}{$F^{I-II}_{E}$}&
     \multicolumn{1}{|c|}{$F^{I-II}_{U}$}& 
     \multicolumn{1}{|c|}{$F^{II-III}_{L}$}&
     \multicolumn{1}{|c|}{$F^{II-III}_{E}$}&
     \multicolumn{1}{|c|}{$F^{II-III}_{U}$}\\
     \hline
     \hline
      $0.125$ &128  &202  &1081  &1052 &42 &16 &0.0 &2.0 &3.2 &3.0 &6.6 &9.0\\
     \hline
      $0.15$  &154  &99   &499   &545  &23 &17 &0.7 &2.0 &3.6 &3.5 &6.5 &9.1\\
     \hline
      $0.2$  &205  &141  &728   &1021 &39 &18 &1.2 &2.0 &3.6 &4.7 &6.3 &8.5\\
     \hline
      $0.25$ &256  &98   &479   &949  &19 &8 &0.0 &1.7 &3.8 &4.4 &5.9 &7.5\\
     \hline
      $0.3$  &307  &83   &364   &878  &13 &6 &1.2 &2.0 &3.9 &4.0 &6.0 &7.3\\
     \hline
      $0.4$ &410  &117  &385   &1169  &23 &7  &   &    &3.7 &4.5 &5.7 &7.0\\
     \hline
      $0.5$  &512  &53   &341   &884  &23 &7 &   &    &     &4.8 &6.0 &7.2\\
     \hline
      $0.6$  &615  &38   &488   &1316  &61 &10 &   &    &     &4.8 &6.1 &7.3\\
     \hline
      $0.7$  &717  &4    &333   &974  &89  &7 &   &    &     &4.9 &5.8 &6.8\\
     \hline
      $0.75$  &768  &2    &256   &960  &108 &7 &   &    &     &4.5 &5.4 &6.6\\ 
     \hline
      $0.8$  &819  &2    &81    &1064 &12  & 5 &   &    &     &    &    &4.3\\
     \hline
\end{tabular}  
\end{center}  
\end{table*}
At each fixed $\lambda$,
we divide the region in $F$ into bin sizes of $\delta F=1$, and count
the number of rule tables of each classes in these bins. From these
results, we estimate the probability densities $P(F_i)$, where $F_i$
is a middle point of that bin.\\
\indent
Let us proceed to the classification of the class II and class III rule
tables.
\subsection{Classification of rule tables in $\frac{1}{4} \le \lambda \le
1-\frac{1}{4}$}
\begin{figure*}
\begin{center}
\scalebox{0.70}{ { \includegraphics{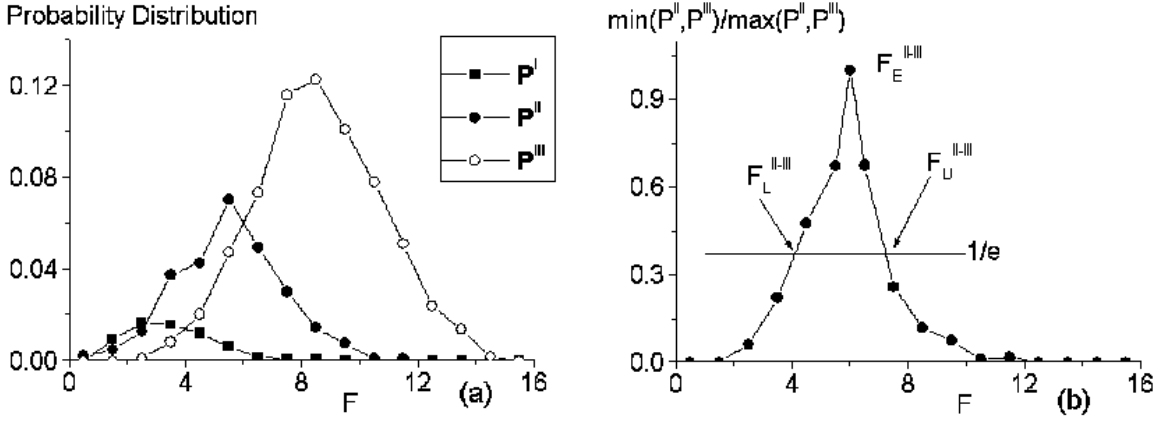} } }
\caption{The probability distributions $P^{I}$, $P^{II}$ and $P^{III}$ and
determination of $F^{II-III}_{E}$, $F^{II-III}_{L}$ and
$F^{II-III}_{U}$ at $\lambda=0.3$.
Fig.~\ref{fig:Phtr_0.3}(a):  $P^{I}$, $P^{II}$ and $P^{III}$
distributions.
Fig.~\ref{fig:Phtr_0.3}(b): An example of the determination of
$F^{II-III}_{E}$, $F^{II-III}_{L}$ and $F^{II-III}_{U}$.}.
\label{fig:Phtr_0.3}
\end{center}
\end{figure*}  
\indent   
The probability distributions of $P^{II}(F)$ and
$P^{III}(F)$ at $\lambda=0.3$ are shown in Fig.~\ref{fig:Phtr_0.3}(a),
and the determination of
the $F^{II-III}_{E}$, $F^{II-III}_{U}$ and $F^{II-III}_{L}$ are
demonstrated in Fig.~\ref{fig:Phtr_0.3}(b). 
For the other $\lambda$ points of Table~\ref{tab:phtr_data},
$F_{E}$, $F_{U}$ and $F_{L}$ are determined in the similar way.
They are summarized in the Table~\ref{tab:phtr_data}.\\
\indent
As already seen in Fig.~\ref{fig:lambda-F-distribution}, the $\lambda$
dependences of the $F^{II-III}_{E}$, $F^{II-III}_{U}$ and
$F^{II-III}_{L}$ are small,
in the region  $0.25 \le \lambda \le 0.75$.
This is also confirmed by the studies at fixed $\lambda$s as shown in
in the Table~\ref{tab:phtr_data}.\\

\subsection{Classification of rule tables in larger and smaller
$\lambda$ region}
\begin{figure*}
\begin{center}
\scalebox{0.65}{ { \includegraphics{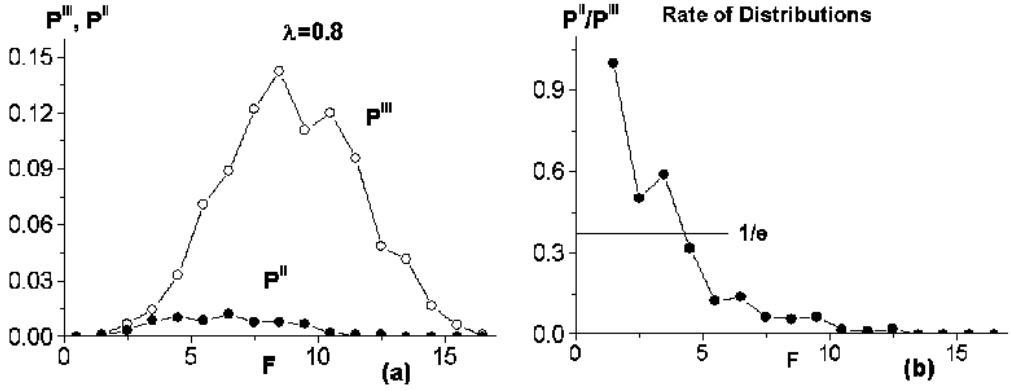} } }
\caption{The probability distributions $P^{II}$ and $P^{III}$
and the determination of $F^{II-III}_{E}$, $F^{II-III}_{L}$ and
$F^{II-III}_{U}$ at $\lambda=0.8$.
Fig.~\ref{fig:L8}(a) shows $P^{II}$ and $P^{III}$. In
Fig.~\ref{fig:L8}(b), it is found that $F^{II-III}_{E}$,
$F^{II-III}_{L}$ could not be obtained.
}
\label{fig:L8}
\end{center}
\end{figure*}

\begin{figure*}
\begin{center}
\scalebox{0.70}{ { \includegraphics{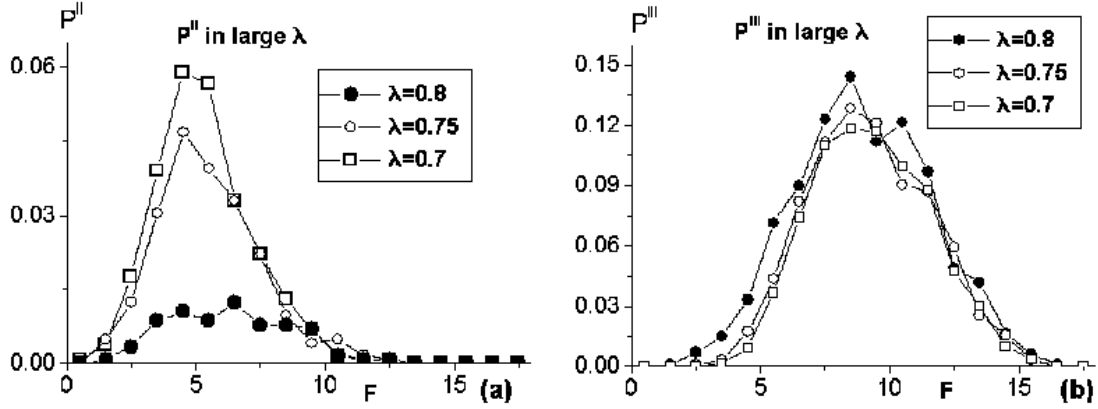} } }
\caption{The $\lambda$ dependences of $P^{II}$ and $P^{III}$ in the region,
$\lambda \geq 0.7$ 
}
\label{fig:Large_L}
\end{center}
\end{figure*}

\begin{figure*}
\begin{center}
\scalebox{0.70}{ { \includegraphics{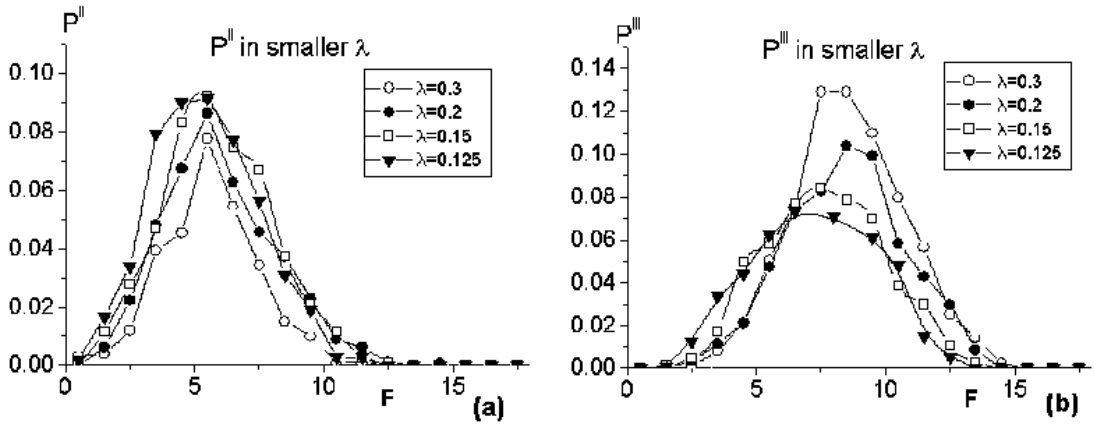} } }
\caption{\label{fig:Small_L}
The $\lambda$ dependences of $P^{II}$ and $P^{III}$ in the
region, $\lambda \leq 0.3$}
\end{center}
\end{figure*}

\begin{figure*}
\begin{center}
\scalebox{0.70}{ { \includegraphics{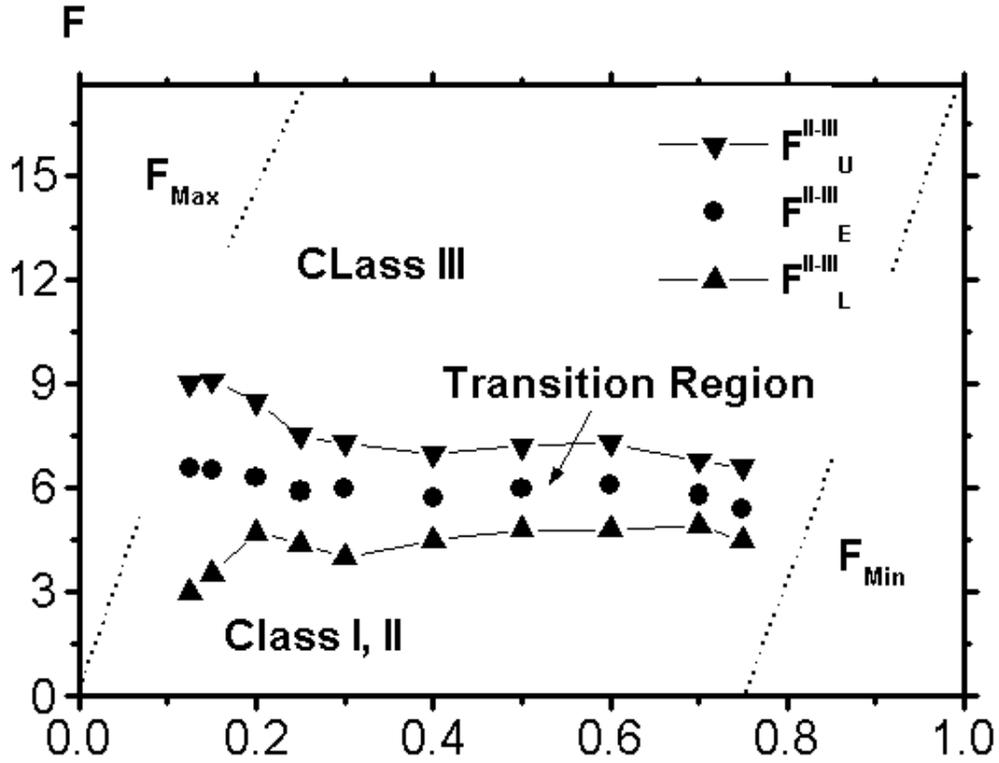} } }
\caption{The phase diagram of CA(5,4).  
The minimum of $F$ ($F_{Min}$) and the maximum of $F$
($F_{Max}$), which are discussed at the end of subsection VI B, are
shown by the dotted lines.
  }
\label{fig:Phase_diagram}
\end{center}
\end{figure*}

The $P^{II}$ and $P^{III}$ at $\lambda=0.8$ are shown in Fig.~\ref{fig:L8}.
It should be noticed that there is no region of $F$ where $P^{II} \geq
P^{III}$. This means that $F^{II-III}_{E}$ and $F^{II-III}_{L}$
disappears. Only $F^{II-III}_{U}$ is determined.\\
\indent
In order to understand what has changed at $\lambda=0.8$, we have
studied the $\lambda$ dependences of $P^{II}$ and $P^{III}$ in the region
$\lambda \geq 0.7$.
They are shown in Fig.~\ref{fig:Large_L}.
It is found that $P^{III}$ gradually increases as $\lambda$
becomes larger but the increase is quite small, while
$P^{II}$ decreases abruptly between $\lambda=0.75$ and
$\lambda=0.8$. As a result $P^{II}$ becomes less than $P^{III}$
in all $F$ regions. This tendency could already be observed in
Fig.~\ref{fig:lambda-F-distribution}(b), but it is quantitatively 
confirmed by the studies at fixed $\lambda$s.\\
\indent
 In the smaller $\lambda$ region ($\lambda \le 0.3$), the behavior of
$P^{II}$ and $P^{III}$ are shown in Fig.~\ref{fig:Small_L}. In this case,
$P^{II}$ is gradually increasing as $\lambda$ decreases but the change is 
small. On the contrary, the decrease in $P^{III}$ is larger.
As a consequence of these changes the transition region of the class II 
and class III pattern classes spread over wider range in $F$.
These results are also summarized in Table~\ref{tab:phtr_data} and shown
in Fig.~\ref{fig:Phase_diagram}.\\
\indent
In the same way, the classification of the class I and class II rule
tables could be carried out. The preliminary results are shown in the
column $F^{I-II}_{E}$, $F^{I-II}_{U}$ and $F^{I-II}_{L}$ of the
Table~\ref{tab:phtr_data}.
 In this case too, it is seen in
Fig.~\ref{fig:lambda-F-distribution}(a), (b) that density of class I rule
tables decreases as $\lambda$ increases, while that of class II rule
tables stays almost constant in $\lambda \leq 0.75$. This feature is
more quantitatively confirmed by the studies at fixed $\lambda$s.
At $\lambda=0.4$, there disappears the region of $F$,
where $P^{I}$ is larger than $P^{II}$ and $F^{I-II}_{L}$ and
$F^{I-II}_{E}$ could not be determined,
just as in the same way as $P^{II}$ and $P^{III}$ distributions at
$\lambda=0.8$. These results are also shown in
Table~\ref{tab:phtr_data}.\\
\indent
However we should like to say that the numbers of the class I 
rule tables, and those of class II rule tables in the region $F \leq 3$ 
are not large. 
Therefore the results may suffer from
large statistical fluctuations. We think that the classification of
class I and class II rule table needs more data to get 
quantitative conclusions, however the qualitative properties will not be
changed.\\
\indent
We proceed to the investigation of the classification of rule tables
outside of these $\lambda$ region.
In $\lambda < 0.25$ region, not all the group 2 rules could be
replaced
by the group 1 rules. Therefore the maximum numbers of group 1
rules $N(g1)_{Max}$ could not become 256, and it decreases to zero as
$\lambda$ approaches to
zero. Then the maximum of $F$, ($F_{Max}$) also decreases to
zero toward $\lambda=0$.\\
\indent
Conversely in $\lambda > 0.75$ region, not all the group 1 rules could
be replaced by the group 2 rules. The minimum of
N(g1) and therefore the minimum of $F$, ($F_{Min}$) could not becomes
0.
The line $F_{Min}$ increases until its maximum at $\lambda=1$.
In Fig.~\ref{fig:Phase_diagram}, we have schematically shown the $F_{Max}$
and $F_{Min}$ lines with dotted lines.
We should like to stress that the dotted line should have some
width due to fluctuations of $N_4$, $N_3$, $N_2$
and $N_1$ caused by the randomness.\\

\section{Discussions and Conclusions}

\subsection{Transmission of initial state informations}
\begin{figure*}
\begin{center}
\scalebox{0.80}{ { \includegraphics{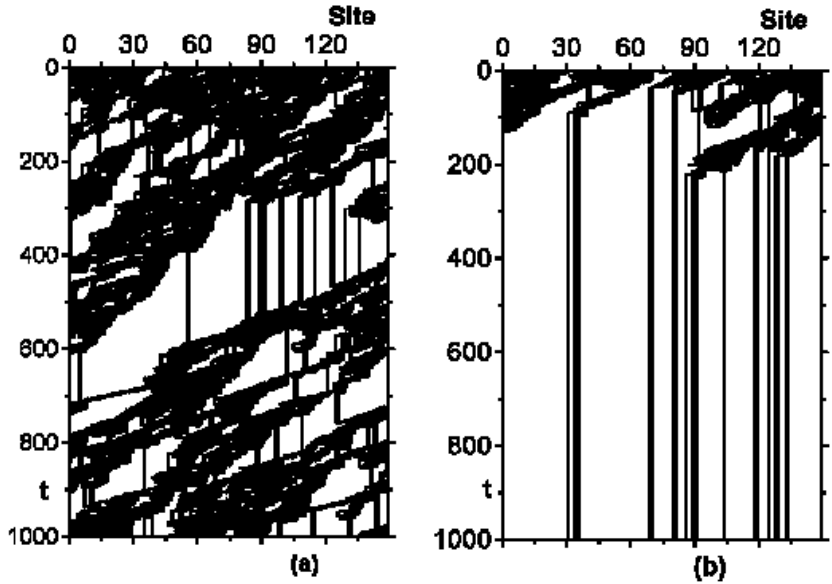} } }
\caption{An example of the transmission of the initial state
information at
$\lambda=0.75$. Fig.~\ref{fig:L75comp}(a), and Fig.~\ref{fig:L75comp}(b) are
generated by the same rule table. Only the difference is the initial
states configurations which are set randomly.}
\label{fig:L75comp}
\end{center}
\end{figure*}

 The computability of the CA is discussed very precisely mainly for
elementary CA (CA(3,2))  in series of
paper from Santa Fe Institute\cite{hanson}.
In this subsection we discuss on the simplest problem of transmission of
the initial state information to the later configurations.\\
\indent
We have found some examples where class II and class IV patterns
appear with similar probability by changing the initial configurations
randomly.
An example is shown in the Fig.~\ref{fig:L75comp}, which is
the transmission of initial state
information to later configurations and is similar to the $\rho=1/2$
problem in the CA(3,2). It is interesting to investigate under what
condition the changes of the pattern classes are taken place. \\
\indent
We have focus on the difference of patten classes between 
class II and class IV, because in this case
differences of the patterns are obvious.
These rule tables are 
found in the wide region in $\lambda$, $0.125 \leq  \lambda \leq
0.8$.  The numbers of the rule tables of this property at fixed
$\lambda$s are also shown
in the column ''Comp'' of Table~\ref{tab:phtr_data}.\\
\indent
In addition,
there are cases where the difference of patters seems
to be realized within the same pattern classes.
In these cases careful studies are 
necessary to distinguish the difference of these patterns.
In this article, we have not studied these cases.
\subsection{Classification of rule table by the intuitive $F$ parameter}
\begin{figure*}
\begin{center}
\scalebox{0.80}{ { \includegraphics{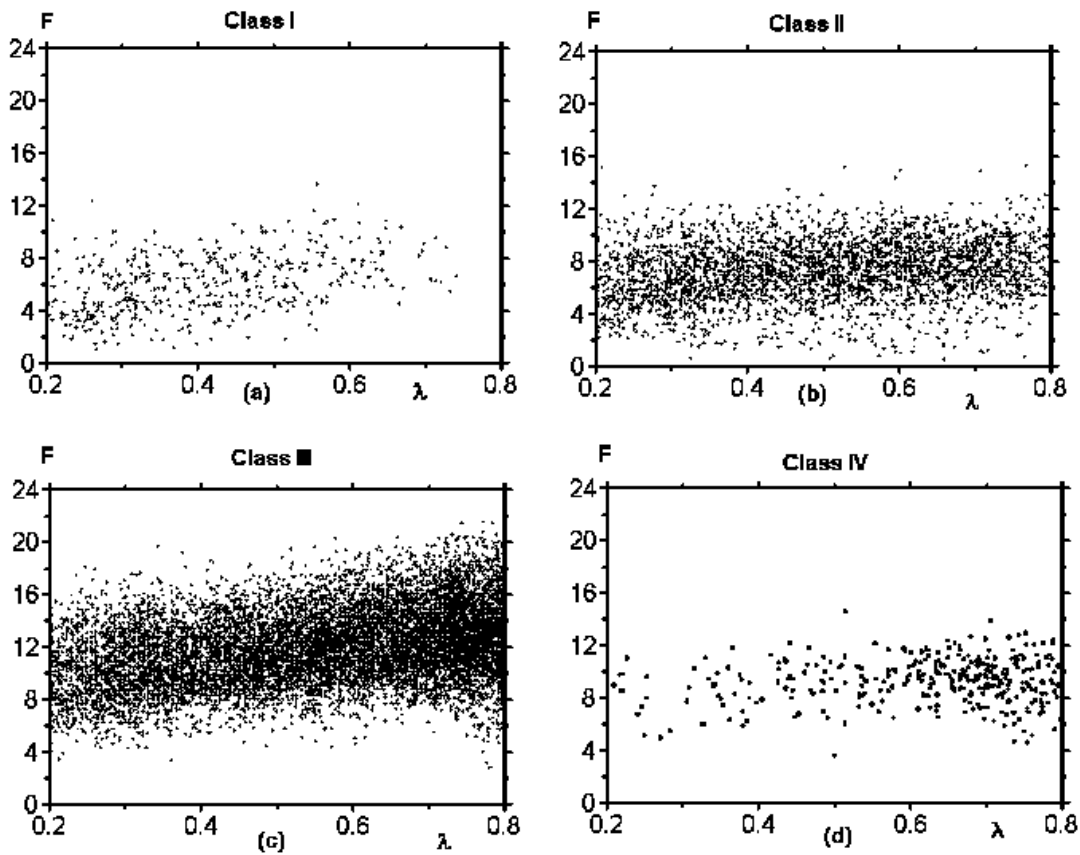} } }
\caption{The distribution of rule tables by using intuitive $F$
parameter. The numbers of the data points are same as in the case of
Fig.\ref{fig:lambda-F-distribution}.
}
\label{fig:intuitive}
\end{center}
\end{figure*}

The methods to determine the coefficients $r_{i}$ in
Eq.~(\ref{eq:taylor}) are not unique.
In the section IV, in order to determine them we have used regression
hyper plane of class IV
rule tables, and in order to obtain 329 class IV rule tables, 
we have generated totally about 14000 rule tables. It is a rather
tedious task.
However optimal set of $r_{i}$ has been close to
the intuitive set of $r_{i}$.\\
\indent
In this subsection we study the classification of rule tables
by the intuitive $F$ parameter.
The same analyses as sections V and VI are carried out and as the similar
figures are obtained in this case too, we will show only the distributions
of the rule tables in ($\lambda$,F) plane in Fig.~\ref{fig:intuitive}.
The Figs.~\ref{fig:intuitive} are very similar to the
Figs.~\ref{fig:lambda-F-distribution}. 
Therefore the classification of the rule table space are almost
same as 
Fig.~\ref{fig:Phase_diagram}, except that the $F_{Max}$
changes from 17.6 to 24.\\ 
\indent
If intuitive $F$ parameter could successfully classify the rule tables
for general
CA(N,K) it would be very convenient, because it reflect the structure of
the CA rules and there is no need to gather a lot of class IV rule
tables, in order to determine $r_i$s. 
Whether it is correct or not must be
concluded after the studies of other CA(N,K)\footnote{The preliminary
studies on the CA(5,3) using the intuitive $F$ parameter show that the
qualitative results are very similar to those of 
CA(5,4). The $\lambda$ dependences of the transition region
are very weak in the region $\frac{1}{3} \leq \lambda \leq
\frac{2}{3}$, and four pattern classes coexist there. The detailed
studies on general CA(N,K) will be reported in the forthcoming
publications.}.  
\subsection{Conclusions and discussions}
We have started to find the mechanism which distinguishes the pattern
classes at same $\lambda$s, and have found that it is closely related
to the structure of
the rule tables. The patten classes of the CA are mainly
controlled by the numbers of the 
group 1 rules, which has been denoted by $N(g1)$.\\
\indent
In the CA(N,K),in the region $\frac{1}{K} \le \lambda \le
1-\frac{1}{K}$,
the maximum of $N(g1)_{Max}$  corresponds to a chaotic limit, 
and its minimum $N(g1)=0$, to a periodic limit.
Therefore in this
$\lambda$ region, we could control the patten classes by changing $N(g1)$
without fail. The method for it is the replacements of
Eq.~(\ref{eq:del_num_RP}).
Using the replacements, we
could study the rule tables which are difficult to obtain by the
''random-table method'' or ''random-walk-through method'' of
Ref.\cite{langton}.
This property could be studied quantitatively by introducing a quiescent
string dominance parameter $F$.\\
\indent
In this article, a quantitative studies are carried out for CA(5,4).
In this case, the group 1 rules are further classified into
5 types as shown in Table~\ref{tab:destruc}, and the classification of
rule tables is carried out in ($\lambda$,F) plane as shown in
Fig.~\ref{fig:Phase_diagram}.
It is seen that the $\lambda$ dependences
of the transition region are very gentle, and
rule tables are classified better by
the $F$ parameter rather than by $\lambda$.
It is interesting whether or not the $\lambda$ dependences of the
transition region  
depend on the models.\\ 
\indent
In the replacement experiments,
we have found the edge of chaos (very long transient lengths) behavior
in many
cases. The examples are shown in  Fig.~\ref{fig:replace1}.
Sometimes they are
observed in some range in $N_3$ or $N_2$. 
This indicates that in 
many cases, the transitions are second-order like. But the widths in the
ranges of $N_3$ or $N_2$ are different from each other, and there are
cases where the widths
are less than one unit in the replacement of RP2 (first-order like).
It is very interesting to investigate under what condition
the transition becomes first-order like or second-order like.
The mechanism of the difference in the transitions
is an open problem and it
may be studied by taking into account effects of group 3 and 4
rules. In these studies another new parameters might be found and a more
quantitative phase diagram might be obtained.\\
\indent
These issues together with finding the  points where the transition
region crosses $F_{Max}$ and $F_{Min}$ lines (dotted lines) in
Fig.~\ref{fig:Phase_diagram},
and the nature of the transition at these points will be addressed 
in the forthcoming publications.

%\clearpage


\begin{thebibliography}{9}
\bibitem{wolfram} S. Wolfram, Physica D 10(1984) 1-35.
\bibitem{wolframs} S. Wolfram, Physica Scripta T9(1985) 170-185.
\bibitem{langton} C.G.Langton, Physica D 42(1990) 12-37.
\bibitem{hanson} J.E.Hanson and J.P.Crutchfield, Physica D
103(1997),169-189, and references therein.
\bibitem{wuensche} A. Wuensche, Complexity Vol.4(1999) 47-66.
\bibitem{binder} P.M. Binder Complex System 7(1993),241-247
\bibitem{oliveira} G.M.B.Oliveira, P.P.B.de Oliveira, Nizam Omar
 Artificial Life 7(2001),277-301
\bibitem{packard}  N.H.Packard, Adaption toward the edge of chaos. In
''Dynamic
Patterns in Complex System''(1988), 293-301, edited by J. A. S. Kelso,
A. J. Mandel,and M. F. Shlesinger,World Scientific,Singapore. 
\bibitem{mitchell} M. Mitchell, J.P.Crutchfield and P.T. Hraber,
Santa Fe Institute Studies in the Science of Complexity, Proceedings
	Volume 19. Reading, MA:, Addison-Wesley.
online paper, http://www.santafe.edu/~mm/
paper-abstracts.html\#dyn-comp-edge.
\bibitem{langton0} C.G.Langton, Physica D 22(1986) 120-149.
\bibitem{langton2} W.LI, N.H.Packard and C.G.Langton, Physica
 D45(1990) 77-94.
\bibitem{suzudo} T. Suzudo, Crystallisation of Two-Dimensional
Cellular Automata, Complexity International, Vol. 6(1999).
on line journal,\\ http://www.csu.edu.au/ci/vol06/suzudo/suzudo.html.
See appendix.   
\end{thebibliography}
\end{document}